# Nonlinear and statistical analysis of ECG signals from Arrhythmia affected cardiac system through the EMD process

Chiranjit Maji[a*], Pratyay Sengupta[b], Anandi Batabyal[b], Hirok Chaudhuri[a]

[a] Department of Physics, National Institute of Technology Durgapur, M G Avenue, Durgapur, West Bengal India

[b] Department of Biotechnology, National Institute of Technology Durgapur, M G Avenue, Durgapur, West Bengal, India

[*]Corresponding author's email id: chiranjit.nitdphysics@yahoo.in
cm.15ph1102@phd.nitdgp.ac.in

## Abstract

The human heart is a complex system exhibiting stochastic nature as reflected in electrocardiogram (ECG) signals. ECG signal is a weak, non-stationary and nonlinear signal, which indicates the health of a heart in terms of temporal variations of electromagnetic pulses from the heart. Abnormal fluctuations in ECG signal invokes the possibility of various cardiovascular disorders, which is diagnosed through intuitive analysis of the ECG reports by the medical practitioners. This could be made fast, accurate and simple by imposing advanced nonlinear tools on the recorded ECG signals. In this paper, a well-known nonlinear technique, i.e., Empirical Mode Decomposition (EMD) method is adopted to extract the hidden information in the recorded ECG signal. Here, we try to explore the human heart as a dynamic model and perform EMD on ECG reports distinguishing arrhythmia from normal data obtained from the widely used MIT-BIH database. EMD essentially involves the decomposition of the signal into a finite number of Intrinsic Mode Functions (IMFs), keeping its original properties unaltered. For analysis, we use the powerful Savitzky-Golay (SG) filter for removing non-stationary noises from the ECG signals. The popular nonlinear parameter Hurst Exponent (H) is estimated for every IMF by R/S technique. We identified a distinct margin of the H of 1st IMFs in between the normal and the arrhythmia affected patients. Our model confirms with 94.92% certainty the chances of occurrence of arrhythmia disease in patients by diagnosing ECG signals without performing other expensive and time-consuming techniques such as Holter test, echocardiogram and stress test.

## 1. Introduction

Recent times have witnessed a widespread prevalence of heart diseases across the globe. According to the World Health Organization (Alwan et al., 2010), cardiovascular diseases are one of the major causes of death worldwide (Alwan, 2011; Rodríguez et al., 2015). A conventional and well-established technique to assess the health of the heart involves the use of electrocardiogram (ECG) signals (Berkaya et al., 2018). Other modern techniques such as Holter test, echocardiogram, and stress test are also used for the diagnosis of a cardiac patient (Mayo_Clinic_HD/2019). However, the availability of such sophisticated and relatively expensive facilities for detection of heart disease are restricted mostly to metropolitan and other big cities of developing countries such India, China which have high population (Gupta et al., 2012). Despite this fact, ECG still plays a vital role towards the cardiologists to understand the health of the heart in an easy and cost-effective way. ECG of an individual is the graphical representation of the electrical potential of the heart, commonly used for detection of the presence of cardiovascular disorder (Subramanian B., 2017). However, cardiologists fail to provide certain vital information about the heart just by conventional ECG-based analysis due to lack of detailed knowledge of proper nonlinear and statistical methods. (de Godoy, M. F., 2016). In this paper, the aim is to establish an algorithm which will help to extract the nonlinear features hidden in ECG signals for easier analysis of these by cardiologists without going much into the details of mathematical tools related to nonlinear methods. Different nonlinear techniques such as multifractal analysis, wavelet transform, recurrence, etc. are often used in earth science, atmospheric science, mathematics, physiology and financial market study, etc. However, Empirical Mode Decomposition (EMD) seems to be an attractive tool among aforesaid methods to efficiently deal with the nonlinearity of a complex system (Huang et al., 1998).

Variations in heart rate are due to systole and diastole induced by the specialized cells of the sinoatrial (SA) node, atrioventricular (AV) node, along with the His-Purkinje system (Draghici and Taylor, 2016). The dysfunction of this system leads to various cardiovascular diseases, including arrhythmia (John and Kumar, 2016; Tse, 2016). Some vital diseases related to the heart comprise coronary artery disease, heart valve disease and congenital heart disease and many more (WebMD_HCD/2016). One major heart ailment is arrhythmia and its symptoms include anxiety, palpitations at rest and exertion, reduced physical ability and breathlessness (Hansson et al., 2004). The occurrence of arrhythmia is also associated with the onset of other severe complications like stroke and heart failure. Based on the rate of heartbeat, arrhythmia has been categorized as (a) Tachycardia (resting heart rate exceeding100 beats per minute) and (b) Bradycardia (resting heart rate below 60 beats per minute) (Mayo_Clinic_AD/2019) whereas heart rate for a healthy person is around 72 beats/min. Efforts have been made so far in the detection of arrhythmia by using diverse adaptive filter structures. Different components of the ECG signal have been individually taken into consideration and compared with normal ECG signals to identify abnormalities such as premature ventricular complexes, atrial fibrillation and others (Thakor and Zhu, 1991). Moreover, emphasis has also been made on the classification of arrhythmia based on RR-interval obtained from ECG signals (Tsipouras et al., 2005). The diagnosis of this disease continues to be largely dependent on intuitive analysis of the medical practitioner based on electrocardiogram (ECG) reports. Interpretation of such reports may vary from one physician to another and may lead to inappropriate treatment by inexperienced physicians. This possesses potent risks for the patients and may lead to fatal outcomes. Moreover, time is a constraint for interpretation of ECG reports of patients suffering from emergency medical issues, which necessitates analysis of these reports within seconds. Moreover, for instant monitoring of emergency medical issues and interpretation of the patient's ECG, the time is a vital factor especially when decisions would be made within seconds (Agarwal et al., 2016). Therefore, an automatic interpretation of medical data like electrocardiogram has been of interest to researchers over the last few decades. Thus, by advancement made in the domain of computational biology, quality of diagnosis of cardiovascular diseases through ECG signals could be improved by the combined effort of the physicists, mathematicians and medical practitioners. In this regard, the tools used in diagnosis would be faster with more efficiency and lower maintenance and most importantly, would be cost-effective (Sahoo et al., 2017; Soorma et al., 2014). Through the last two decades, techniques based on statistics have been applied to figure out the dynamical features, namely nonlinearity, of the system of interest (Mukherjee, 2012; Shumway and Stoffer, 2011; Chatfield, 2004). Among various data types, medical data such as the electrocardiogram (ECG) and electroencephalogram (EEG) data, Parkinson's disease data etc. exhibit a significant amount of nonlinearity (Yao et al., 2009). Studies in this area include the nonlinear technique such as bi-spectral analysis of ECG signals of patients suffering from atrial fibrillation (AF) as well as that detected during different stages of sleep (Sezgin, 2013; Fell et al., 2000).

Our study is focused on minimizing the aforesaid difficulties by developing novel computational methods to identify patients with arrhythmia by adopting EMD techniques on ECG signals. In view of this, we have collected ECG data of patients suffering from arrhythmia and normal ECG data of healthy persons from the official website of PhysioNet (PhysioNet). Applications of computational biology and data sciences on various biological complex systems greatly assist us in understanding many physiological processes. Diverse techniques such as multifractal analysis, wavelet transform, recurrence, etc. are often used in physiology and medical sciences. EMD seems to be an attractive tool among aforesaid methods. EMD essentially involves the decomposition of the signal into a finite number of IMFs keeping its original properties intact. Here, IMFs of each data series (disease as well as normal) have been evaluated by means of applying EMD and the Hurst exponent of every IMF has been computed. Our goal is to distinguish arrhythmia affected patients and normal persons based on a comparative study of Hurst exponents of a particular IMF retrieved from the ECG signals. Our attempt to extract information pertaining to the health of the heart using EMD modulated ECG signal may assist to establish a new feature in cardiology science.

## 2. Materials and Methods

### 2.1. ECG data acquisition

With the purpose of distinguishing a person suffering from arrhythmia from a healthy (normal) individual, full-length ECG time series data of 48 arrhythmic patients has been collected from popular MIT-BIH Arrhythmia Database (mitdb) (PhysioNet/MIT-BIH_AD/2018). The data sets include ECG time series data of 24 men aged between 32 to 89 years and of 22 women aged between 23 to 89 years (Moody and Mark, 2001; Goldberger et al., 2000). On the other hand, the normal data (healthy person) has been collected from MIT-BIH Normal Sinus Rhythm Database (nsrdb) (PhysioNet/MIT-BIH_NSRD/2018) consisting of 18 long-term ECG signals of subjects having no significant arrhythmia. The subjects include 5 men and 13 women, aged between 26 to 45 and 20 to 50 respectively (Goldberger et al., 2000). The signal used here for the analysis is a modified limb lead II (MLII), obtained by placing the electrodes on the chest of the patients. We did not include the records of two patients with patient IDs 102 and 104 from MIT-BIH Arrhythmia Database as the required MLII data were not available due to surgical dressings on the above-mentioned patients. On the other hand, we also consider the data from ECG1 mode, which are ECG signals (Normal Sinus Rhythm Database) relating to healthy persons and these data sets are regarded as complementary to MLII data of the arrhythmia database. Here, total no. of data points of each of the disease data series is 21600 with frequency 360.01 $sec^{-1}$ whereas the same for the normal data series is 7680 with frequency 128 $sec^{-1}$. Therefore, each of the data series is recorded for 60 sec time duration. Moreover, the variation of the ECG signals of the arrhythmic patients is bounded in between -3.105 and 2.8 mV while it ranges from -2.215 to 3.675 mV in case of normal patients.

### 2.2. Data smoothening by Savitzky–Golay filter

A filter is utilized for processing of signals in order to selectively isolate a particular frequency or range of frequencies from an assortment of multiple frequencies in a signal (Jagtap & Uplane, 2013). The choice of appropriate filter for processing of the system generated signals requires maximum noise reduction with minimal signal distortion. Thus, extensive research is being carried out on choosing the digital filter for de-noising purpose, and it helps to improve the results significantly (Yadav et al., 2015). Several methods are used to filter the noise during real-time biomedical data preprocessing (Agarwal et al., 2016). One of the easiest filters for smoothing the nonlinear time series is moving average. In moving average, a fixed number of points, called frame, are taken and their ordinates are added followed by division by the number of points to get the average ordinate. Then, the point at one end of the frame is dropped and the next point at the other end is added. This process is performed repeatedly until the endpoint of time series is attained. But, this method is sometimes clearly undesirable due to the accompaniment of the degradation of peak intensities (Savitzky and Golay, 1964). Thus, one of the best filters for noise clearing of biomedical data, including ECG signals, seems to be Savitzky–Golay (SG) filter (Hargittai, 2005; Savitzky and Golay, 1964). The fundamental principle of SG filter is to consider $2n + 1$ equidistant points ($-n, \ldots, 0, \ldots, +n$) taking n = 0 as a center to represent a polynomial of degree $p$ (where $p$ is less than $2n + 1$). A set of points is to be fitted to some curve. For this purpose, SG filter computes the value of the least square polynomial (or its derivative) at a point, i = 0, over the decided frame range. This filter applies the method of linear least squares for data smoothing, which helps to maintain the original shape of the signal (Sahoo et al., 2017) (Supplementary figure: Fig. S1). It is noteworthy that increasing the order and decreasing the frame size of SG filter leads to better conservation of short fluctuations while needless use of high order polynomial with extremely short frame size may lead to over-fitting (Yadav et al., 2015). A SG filter generally requires pre-determined values of order and frame depending on the frequency and length of the data. Usually, trial and error method or prior experience is required to decide the satisfactory values of parameters. The values of such parameters are considered in our study and other necessary information of the data sets are also included as shown in Table I.

Table I: Values of different parameters considered for the SG filter of time series data

| Choice of order and frame of SG filter | Disease | Normal |
|---|---|---|
| Time period (sec) | 0.0027777 | 0.0078125 |
| Length of data | 21600 | 7680 |
| Frame of SG filter | 37 | 13 |
| Order of SG filter | 3 | 3 |

## 2.3. Adaptation of EMD and extraction of IMFs

EMD is a method developed by N. Huang, commonly used to analyze phenomena having nonlinear and non-stationary characteristics (Huang et al., 1998). This method involves decomposing the original signal into a finite number of components, slowly varying amplitude and phase, termed as Intrinsic Mode Functions (IMFs). The IMFs represent the oscillatory portion of the signal (Soorma et al., 2014). EMD is an adaptive method due to the production of signal specific IMFs, which are of a very close to the original signal components visually, with utterly indistinguishable frequencies and amplitudes (Tolwinski, 2007). This multi-scale analysis method, which is used widely for the prediction of the signal's trend, does not rely on any prior knowledge (Blanco-Velasco et al., 2008; Tolwinski, 2007). It is a data-driven mechanism whereas other well-known nonlinear data analysis methods, like Fourier and wavelet-based methods, need some pre-defined basis functions for representing a signal (Blanco-Velasco et al., 2008). The IMF obtaining process from the signal is called the sifting process (Barman et al., 2016; Davis, 2012; Nunes & Deléchelle, 2009; Blanco-Velasco et al., 2008; Peng et al., 2005; Nunes et al., 2003). The entire process is described as follows:

1. x(t) is considered as a signal, and the initiation of the sifting process is started by the identification of all the local maxima and minima of x(t).
2. Then all the local maxima are to be connected by a cubic spline curve using interpolation, said as upper envelope $e_u(t)$. Similarly, the local minima are connected to get the lower envelope $e_l(t)$.
3. The mean of the upper envelope and the lower envelope is calculated and denoted by $m_1(t)$, i.e.,

$$m_1(t) = \frac{[e_u(t)+e_l(t)]}{2} \quad (1)$$

4. The mean is then subtracted from the main signal x(t) to get the first proto-IMF $h_1(t)$. Thus,

$$\mathrm{h}_1(\mathrm{t}) = \mathrm{x}(\mathrm{t}) - \mathrm{m}_1(\mathrm{t}) \quad (2)$$

5. Due to multiple extrema present in between two consecutive zero crossings, the sifting process is to be applied continuously to $\mathrm{h}_\mathrm{k}(\mathrm{t})$, the $\mathrm{k}^{\mathrm{th}}$ proto-IMF. Once, this satisfies the IMF conditions, the first IMF $\mathrm{c}_1(\mathrm{t})$ is obtained.
6. The sifting process gets stopped once reaching the stopping criterion which is characterized by the Sum of Deviations (SD) as

$$SD = \sum_{t=0}^{T} \frac{|h_{k-1}(t)-h_k(t)|^2}{h_{k-1}^2(t)} \quad (3)$$

The first IMF $c_1(t)$ is obtained when the SD is smaller than or equal to the threshold value $SD_{max}$. The typical values of $SD_{max}$ lie between 0.2 and 0.3.

7. After that, the IMF is deduced from the original signal to find the first residual signal $r_1(t)$ by

$$r_1(t) = x_1(t) - c_1(t) \quad (4)$$

8. This residual signal is considered as the original signal to produce further a pair of IMF and residual signals. The procedure is allowed to continue until the $N^{th}$ residue $r_N(t)$ turns to be a constant or with a single extremum or having a monotonic slope.

9. Finally, combining all the above steps, the original signal can be expressed as

$$x(t) = \sum_{n=1}^{N} c_n(t) + r_N(t) \quad (5)$$

Each IMF should satisfy the following characteristics (Ren et al., 2014) as described below:

A. The number of zero-crossing must be equal or differ by one unit to the number of extrema (assuming it has at least two extrema).

B. Each IMF should be symmetrical with respect to the local mean.

## 2.4. Determination of significant IMFs

After getting all IMFs, the significant one is to be chosen (Barman et al., 2016; Peng et al., 2005) from them. For this purpose, the correlation ($C_{xc_n}$) between the IMFs and the corresponding original time series data are calculated using the following equation (Wharton et al., 2013; Jha et al., 2006).

$$C_{xc_n} = \frac{\int x(t)c_n(t)}{\sqrt{\int x^2(t)dt \int c_n^2(t)dt}} \quad (6)$$

A hard threshold $\lambda$ is used to select the significant IMFs. $\lambda$ is a ratio of maximal $C_{xc_n}$ described as

$$\lambda = \frac{\max(C_{xc_n})}{\eta} \quad (7)$$

Where $\eta$ is a ratio factor greater than 1.0 (Peng et al., 2005). In our study, $\eta = 25$ is considered as per our convenience following the trial and error method. IMFs with $C_{xc_N} \geq \lambda$ were considered to be significant. MATLAB R2016a was used to calculate the correlation coefficients.

## 2.5. Estimation of Hurst exponents

Hurst Exponent (H) is one of the important factors to enumerate the measure of irregularity existing in a time series. This exponent was introduced by popular hydrologist H. E. Hurst in 1951, during dealing with the reservoir control problems of dams on the Nile river (Granero et al., 2008; Hurst, 1951;). Mandelbrot and Wallis later explained this methodology to be superior to the autocorrelation, ANOVA as well as to the spectral analysis (Mandelbrot and Wallis, 1969). Whether the value of H greater or lesser than 0.5 indicates the pattern of the nonlinearity of the data set. When the value of H is 0.5, it is called a random walk (Zhang et al., 2008). The process then follows the Brownian motion (Granero et al., 2008). If the value of the exponent is greater than 0.5, it suggests a persistent pattern, i.e., the time series have a long-term positive correlation, and it will be more regular in preserving its trends. On the contrary, the value of H less than 0.5 indicates an anti-persistent pattern. In the extremities, like when H→0, data usually follows no pattern and alter unpredictably, irregularly, exhibiting white noise features. On the other hand, H=1 refers to the predictable nature of a data set (Hurst, 1951; Zhang et al., 2008).

When the functional contrast among individual IMFs cannot be delineated using basic statistics and patterns in the noisy high-frequency data, the H might be a powerful measure for the characterization of patients (Feng and Vidakovic, 2015). In this paper, we introduce the novel approach of using the H for feature extraction of all IMFs obtained from ECG reports of arrhythmic patients as well as of normal subjects. Several techniques are available in time series analysis to figure out H from a time series data (Resta, 2012). One of the best methods to estimate H is R/S technique, proposed by Mandelbrot and Wallis in 1969 (Mandelbrot and Wallis, 1969). The Hurst exponent, the self-similarity parameter, provides the power of long-range dependence in the IMFs. The steps to get H using R/S method are described below

1. The analysis begins by dividing the IMFs of length L into d sub series ($Z_{i,m}$) of length n.
2. The sample mean ($E_m$) and the standard deviation ($S_m$) are calculated for each sub-series, m = 1,….,d
3. Then the sub-series data ($Z_{i,m}$) is normalised as

$$X_{i,m} = Z_{i,m} - E_m \text{ for } i = 1,\ldots, n. \quad (8)$$

4. A cumulative time series is created as

$$Y_{i,m} = \sum_{j=1}^{i} X_{j,m} \text{ for } i = 1,\ldots, n. \quad (9)$$

5. The range is calculated from the cumulative series as

$$R_m = \max\{Y_{1,m}, \ldots, Y_{n,m}\} - \min\{Y_{1,m}, \ldots, Y_{n,m}\}. \quad (10)$$

6. The range is then re-scaled by dividing by $S_m$.
7. Finally, for all the sub-series of length n, the mean value of the re-scaled range is considered as follow

$$\left(\frac{R}{S}\right)_n = \frac{1}{d}\sum_{m=1}^{M}\frac{R_m}{S_m} \quad (11)$$

8. R/S statistics asymptotically follows the relation $(R/S)_n \approx cn^H$. The value of H can be evaluated by doing simple linear regression with the help of the equation

$$\log(R/S)_n = \log c + H\log n \quad (12)$$

## 2.6. Quantitative study of Hurst exponents

The mean and median of H of IMFs were calculated to explain the experimental result. We performed a two-tailed student's t-test with unequal variances ($p < 0.05$) to identify the significant difference in H of 1st IMFs of arrhythmic patients from the normal ones. A p-value refers to the probability of any event that occurred by chance due to experimental fault. A p-value of 0.05 or below indicates the data to be significant (Dahiru, 2008).

## 2.7. Subgrouping of the dataset based on age and gender

The data sets were grouped into two subgroups based on age and gender. The age-based group consists of four subgroups of below 30, 30-50, 50-70, and above 70. While the gender-based groups were separated for male and female individuals. The above mentioned quantitative procedures (see 2.6) were also followed for each subgroup.

## 2.8. Calculation of correlation coefficient- 'Pearson product moment (r)'

In this paper, an attempt has been made to delineate the degree of association between the ages of subjects (both diseased and normal) and the corresponding values of H of $1^{st}$ IMFs by utilizing statistical approaches. For the purpose of analysis, two distinct recorded data sets, i.e. two variables namely, (a) ages of diseased and normal individuals and (b) corresponding H values of $1^{st}$ IMFs were taken into consideration. Correlation coefficients are calculated by means of estimating the most common correlation coefficient: 'Pearson product moment' which is defined as follows (Zaiontz, 2017):

$$r = \text{cov}(x, y)/\sigma(x)\,\sigma(y) \quad (4)$$

$$\text{and, } cov(x, y) = \sum_{i=1}^{N}(x_i - \bar{x})\,(y_i - y)/(N-1) \quad (5)$$

where, i= 1, 2..., N and N is the length of the data set;

$x_i$ (or $y_i$)= The $i^{th}$ value of x (or y).

$\bar{x}$ and $\bar{y}$, are the mean and σ (x), σ (y) are the standard deviations of the data sets x and y, respectively.

Important statistical tools that are frequently employed to decipher the linear relationship between two variables include correlation coefficient (r). By definition, the two variables are said to be perfectly linear as well as positive or correlated to some extent positively, if the value of ‘r’ is +1 or any positive fraction, i.e., high values of x are linked with higher values of y and vice versa. On the other hand, if value of ‘r’ is -1 or any negative fraction, the two variables are said to be perfectly linear as well as negative or correlated to some extent negatively, i.e., high values of x are associated with low values of y and vice versa. r=0 refers the two variables to be uncorrelated i.e. variables are random in nature (Chaudhuri et al., 2019). Moreover, apart from mathematical approach, the geometrical relation between two variables x and y can be graphically designed through scatter plots along with the confidence ellipse. In general, the shape of the confidence ellipse with scatter plot invokes the possible nature of correlation. For example, random data have a much more tendency towards a circular shape and positively/or negatively correlated data are represented by narrower shaped ellipse. The value of the correlation coefficient (r) is defined by the slop of the major axis of the ellipse. Positive, negative and zero value of the slope indicate positive, negative nature of correlation and uncorrelated data respectively (Chaudhuri et al., 2019).

The whole methodology is followed by the schematic diagram, as shown in Fig.1. MATLAB R2016a was used for filtering, EMD, and estimating H. Other statistical analyses were carried out using OriginPro 8.5, GraphPad Prism v7, MS Excel, etc.

## 3. Results and Discussions

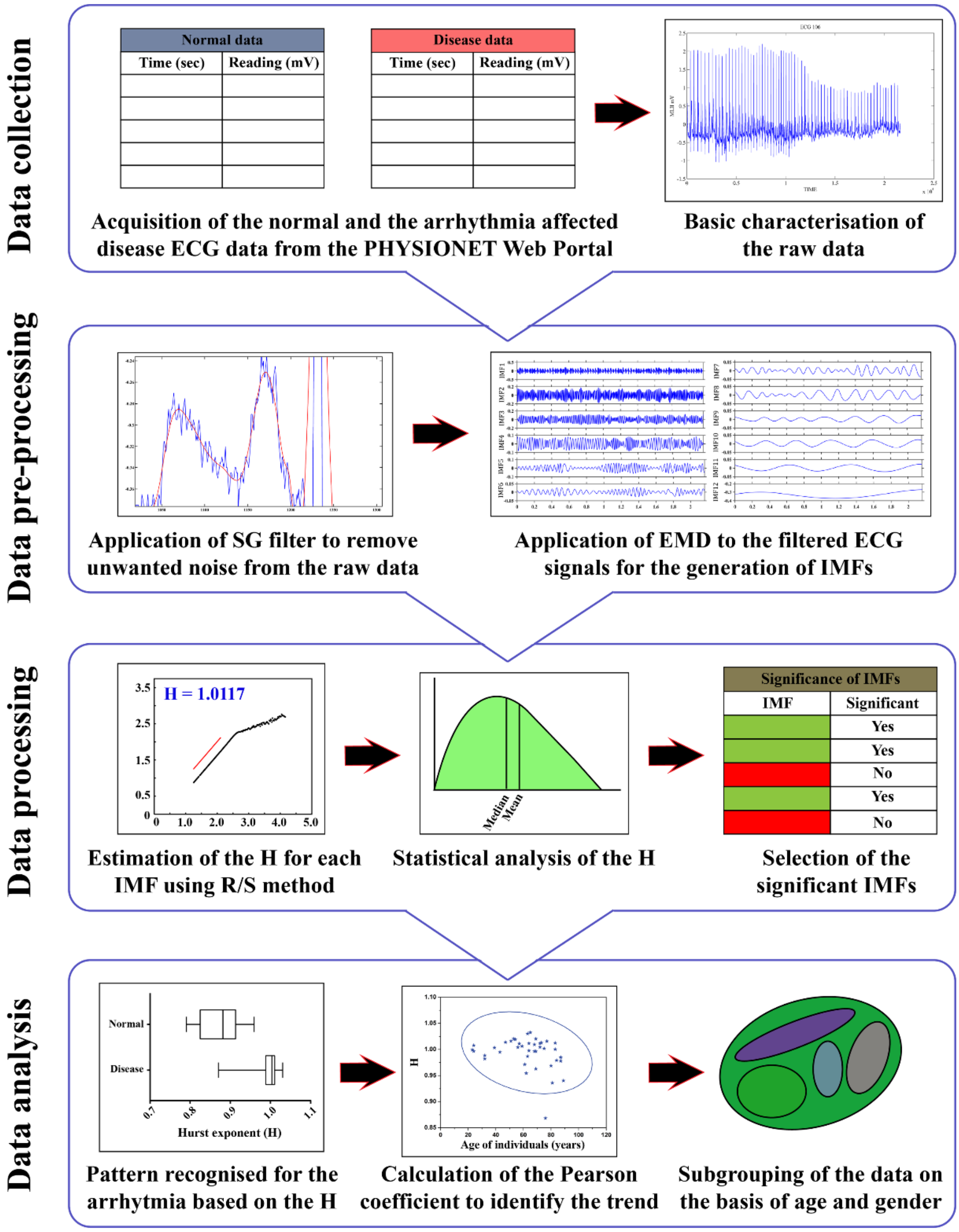

Figure 1: Flow chart of the methodology followed in the study

SG filter was applied to the time series data of 46 arrhythmia-affected patients (called "disease data" here) and 18 normal patients to eliminate the noise present in the data sets. The temporal variation of the smoothened data sets of one normal patient (Patient ID: 19090) and one diseased patient (Patient ID: 100) are plotted as shown in Fig. 2. The figure reflects the distinction between a normal and diseased patient's electrocardiogram. Here, the normal ECG contains 47 peaks in 60 seconds, whereas 74 peaks are present in that of the disease data in the same interval. Furthermore, every cycle (wavelength) is repeated similarly in the normal ECG data, but that is not observed in the disease data sets. Values of the maximum (and minimum) amplitude of all cycles are nearly constant throughout the measurement time for the normal data. However, a rapid variation on the same is observed in the ECG of disease data. Therefore, from the primary observation on the ECG plot in Fig. 2, one can understand the deviation of

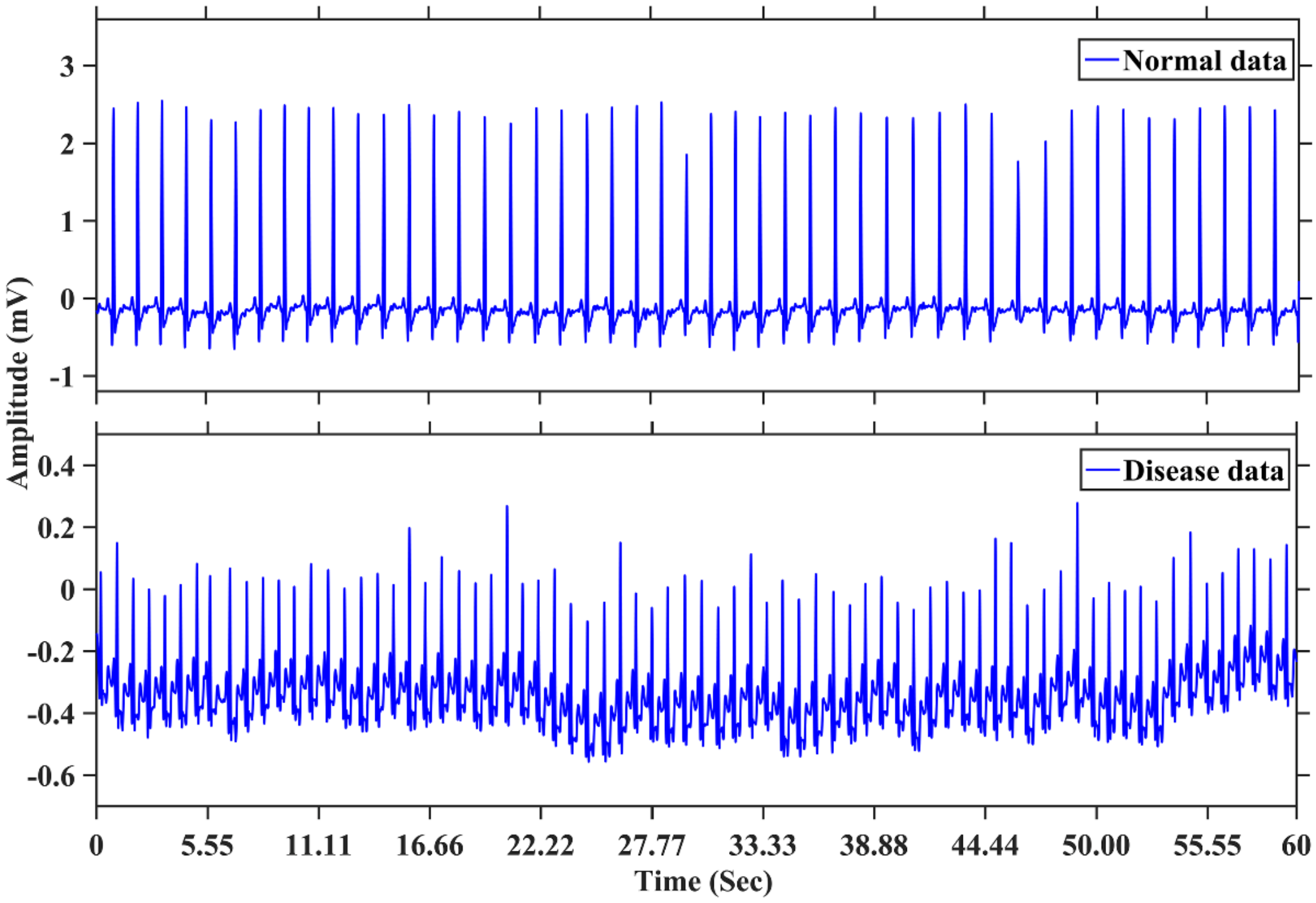

Figure 2: Temporal variation of the ECG data of one normal (Patient ID: 19090) and one diseased (Patient ID: 100) patient

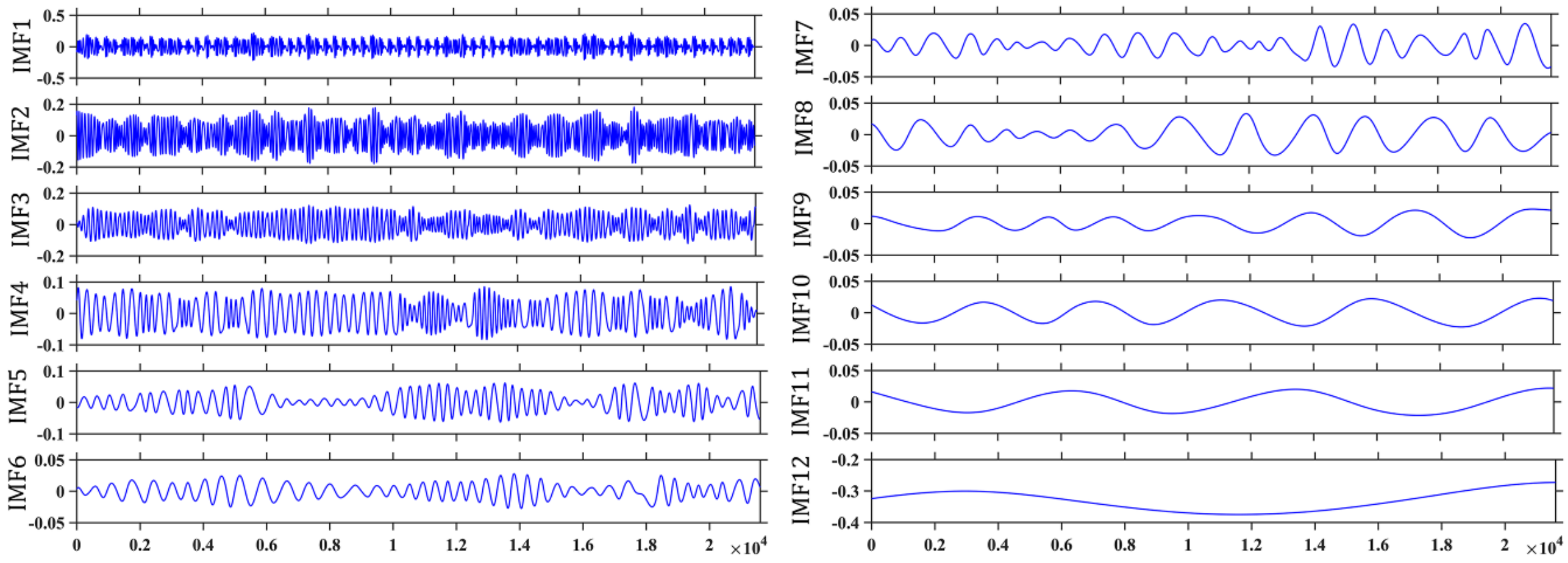

Figure 3: The plots of all IMFs corresponding to one diseased patient's data (Patient ID: 100)

ECG of a diseased patient from that of a normal patient. However, that does not quantify the seriousness and the type of the disease. Therefore, the EMD analysis was applied to each of the data series (total 64 series, Supplementary table: Table SI) following the sifting process as described earlier to retrieve the IMFs. Here, the maximum number of iteration used is 150 to get the maximum number of IMFs. The number of IMFs retrieved from all the series varies between 10 and 14. The IMFs are also formed as a new time series with the data points equal to that of the original time series. Moreover, the time series of IMFs corresponding to one disease data (Patient ID: 100) are plotted as shown in Fig. 3. Now, H of all the IMFs of all the time series (Supplementary table: Table SII) were evaluated (Fig. 4) to enumerate (a) regularity, (b) scaling in the data series. Comparison of H of different IMFs for a normal

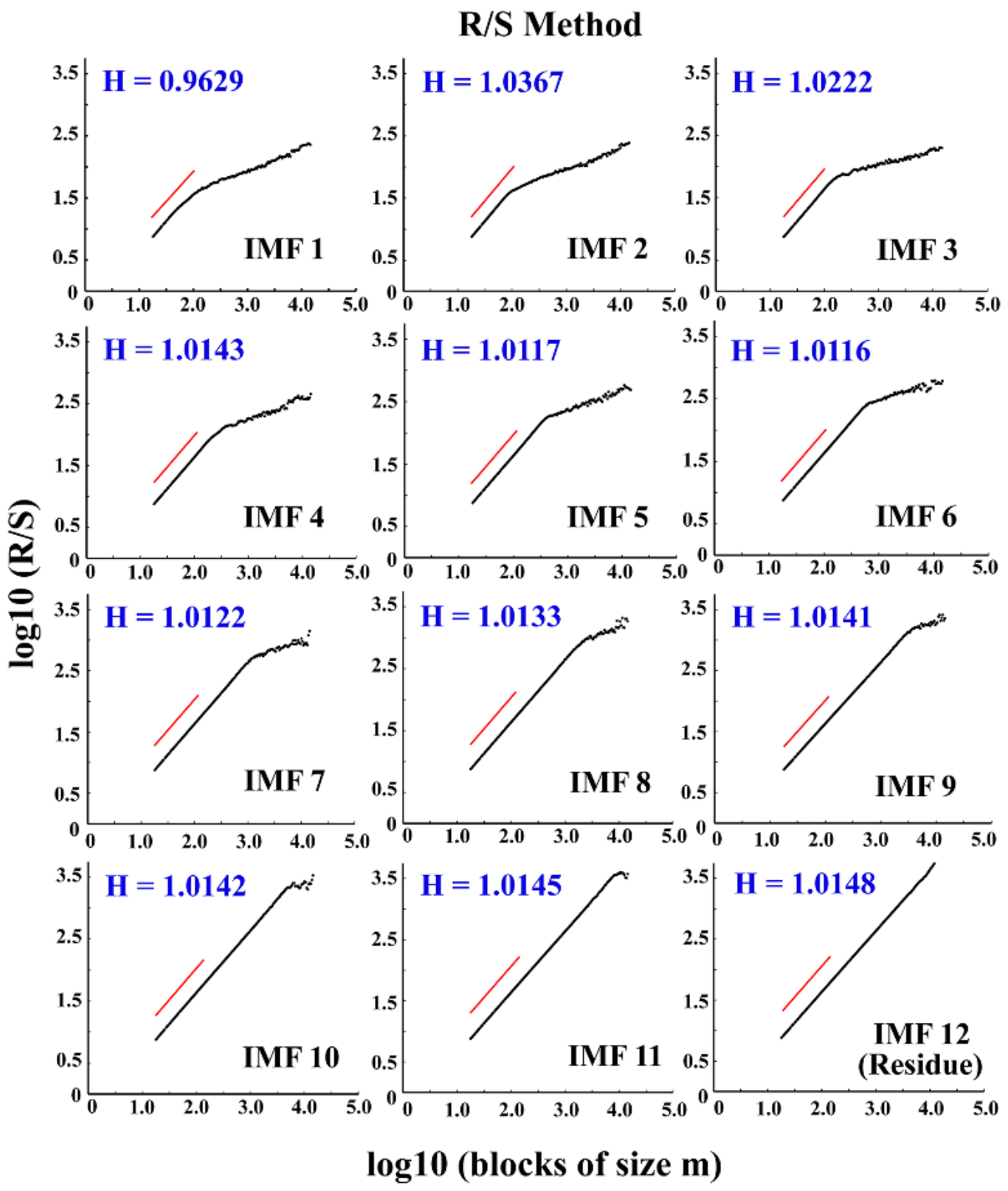

Figure 4: Estimation of H of all IMFs of a diseased patient's data (Patient ID: 100) by R/S technique

Table II: Comparison of H of different IMFs for a normal and a diseased data

| | H of different IMFs | | | | | | | | | | | |
|---|---|---|---|---|---|---|---|---|---|---|---|---|
| Patient ID | IMF 1 | IMF 2 | IMF 3 | IMF 4 | IMF 5 | IMF 6 | IMF 7 | IMF 8 | IMF 9 | IMF 10 | IMF 11 | IMF 12 |
| 19090 (Normal data) | 0.8766 | 1.0164 | 1.0449 | 1.0348 | 1.0192 | 1.0180 | 1.0150 | 1.0160 | 1.0163 | 1.0168 | 1.0179 | 1.0181 |
| 100 (Disease data) | 0.9630 | 1.0368 | 1.0223 | 1.0144 | 1.0117 | 1.0117 | 1.0122 | 1.0134 | 1.0142 | 1.0143 | 1.0145 | 1.0148 |

(Patient ID: 19090) and a disease (Patient ID: 100) data are presented in Table II. As the signature of the arrhythmia with respect to the H was observed in the case of the $1^{st}$ IMFs only, thus each such series were examined whether that was significant or not with the help of the equation no. 6 and 7 mentioned in methods (Supplementary table: Table SIII). Here, $1^{st}$ IMFs corresponding to the patients (IDs) 116, 122, 210, 219, and 223 (diseased patients) were observed to be insignificant and eliminated from further analysis. Hence, the analysis was performed with 59 data series. The H of all the 1st IMFs of all the time series are considered here and are listed in Table III. The p-value enumerated by two-tailed student's t-test between the H of all 1st IMFs corresponding to the disease data and normal data is $0.3\times10^{-9}$ ($p < 0.001$), which indicates that the difference in H related to the individual groups is significant. From the Table III, it is notable that the H of all the 1st IMFs of the time series corresponding to normal patients vary from 0.7886 to 0.9613 with mean value 0.8728 and median 0.8816 whereas the same for the diseased patients are seen to fluctuates between 0.8685 to 1.0324 with mean 0.9940 and median 1.0018. If we neglect the H's values for the patients (ID) 19093, 19140 and 232 in our analysis, the variations changed to 0.7886 to 0.9297 (with mean 0.8631 and median 0.8705) for normal patients and 0.9361 to 1.0324 (with mean 0.9972 and median 1.0025) for diseased patients. Therefore, there is a distinct boundary between the H values of normal and diseased patients (Fig. 5). It implies that for normal patients, H is always less than 0.9300 and that is greater than the same corresponding to diseased patients except for the above mentioned three patients. The fact implies that the dynamical behavior of the cardiac system of a normal person differs noticeably from the same of an arrhythmia affected individual. Therefore, the statement is true for 56-time series data except 03, i.e., there is a probability of 94.92% towards the predictability of arrhythmia disease in a patient by this analysis procedure.

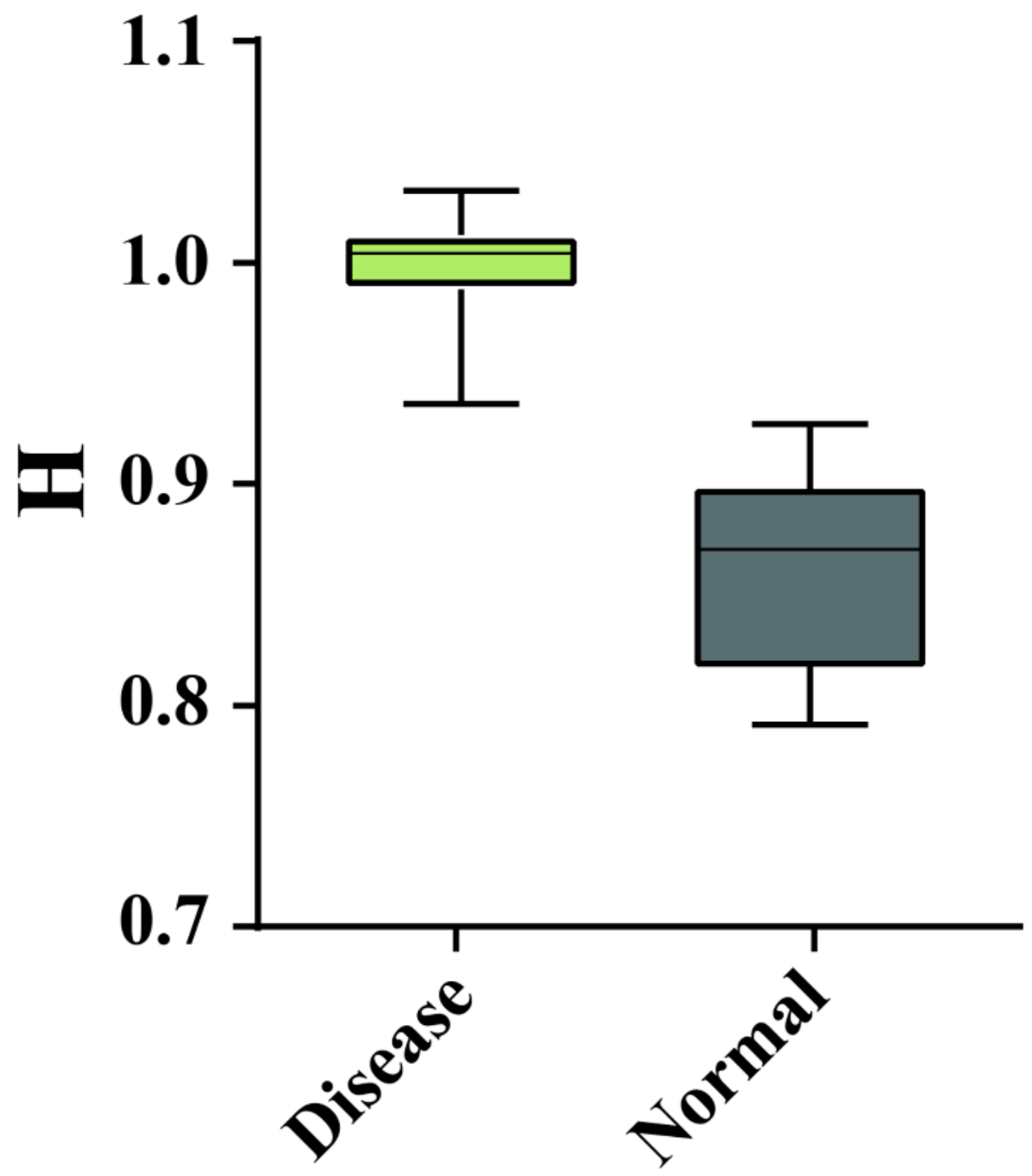

Figure 5: Box and whisker plot for H of all 1st IMFs of diseased and normal patients

Table III: Values of H of 1st IMFs of each (normal and arrhythmia disease) patients' data

| Normal data | | | | | | | | | | | |
|---|---|---|---|---|---|---|---|---|---|---|---|
| **Patient ID** | 19090 | 19093 | 19140 | 19830 | 16265 | 16272 | 16273 | 16420 | 16483 | 16539 | 16773 |
| **H value** | 0.8766 | 0.8436 | 0.9022 | 0.8644 | 0.8876 | 0.8605 | 0.8002 | 0.8073 | 0.9138 | 0.8261 | 0.8866 |
| **Patient ID** | 16786 | 16795 | 17052 | 17453 | 18177 | 18184 | 19088 | | | | |
| **H value** | 0.8126 | 0.7886 | 0.9297 | 0.9185 | 0.9399 | 0.9613 | 0.8915 | | | | |
| **Diseased data** | | | | | | | | | | | |
| **Patient ID** | 100 | 101 | 103 | 105 | 106 | 107 | 108 | 109 | 111 | 112 | 113 |
| **H value** | 0.9630 | 1.0018 | 1.0072 | 1.0129 | 0.9966 | 1.0303 | 0.9779 | 1.0115 | 1.0137 | 1.0189 | 1.0077 |
| **Patient ID** | 114 | 115 | 117 | 118 | 119 | 121 | 123 | 124 | 200 | 201 | 202 |
| **H value** | 1.0044 | 1.0033 | 0.9964 | 1.0202 | 1.0161 | 1.0124 | 0.9869 | 1.0157 | 0.9975 | 0.9965 | 1.0096 |
| **Patient ID** | 203 | 205 | 207 | 208 | 209 | 212 | 213 | 214 | 215 | 217 | 220 |
| **H value** | 0.9950 | 1.0061 | 0.9405 | 0.9994 | 0.9544 | 0.9821 | 0.9714 | 1.0203 | 0.9361 | 1.0324 | 0.9847 |
| **Patient ID** | 221 | 222 | 228 | 230 | 231 | 232 | 233 | 234 | | | |
| **H value** | 1.0004 | 0.9792 | 0.9687 | 0.9885 | 1.0115 | 0.8685 | 1.0110 | 1.0042 | | | |

Pearson product moments (r) between the H values of 1st IMFs (for diseased as well as normal individuals) and the corresponding age of the individuals were calculated to verify the linear relationship between the above-mentioned variables. The value of 'r' for the normal patients was evaluated to be positive (0.2671) whereas the said value was negative (-0.2645) for the diseased patients. Therefore, a positive (upto some extent) correlation is present for the normal individual and negative (upto some extent) correlation is reflected for the diseased patients. The scatter plots along with the confidence ellipse for the normal and disease cases are illustrated in Fig. 6. Hence, the result also strengthens the afore-said conclusion. The disease and normal patient's data could be also distinguished in terms of the 'r' values.

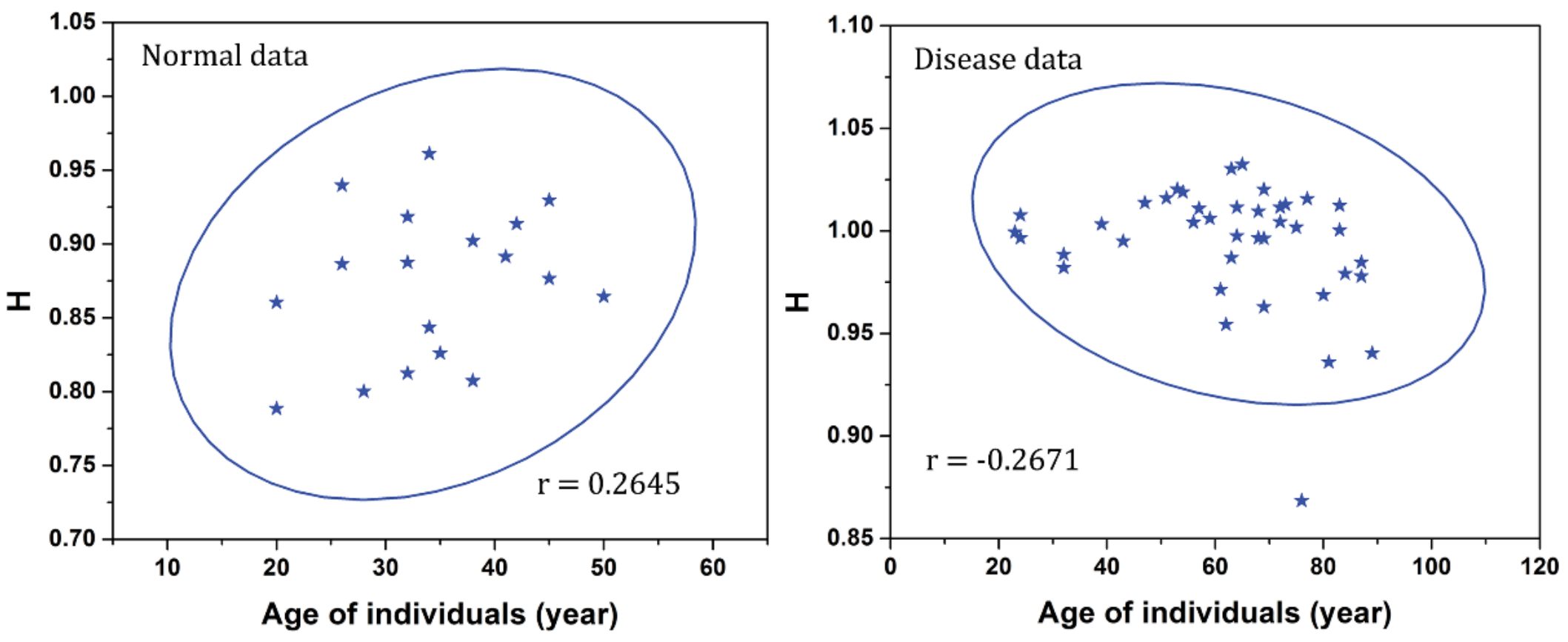

Figure 6: Scattered plot along with confidence ellipse for H of all 1st IMFs of normal and diseased patients

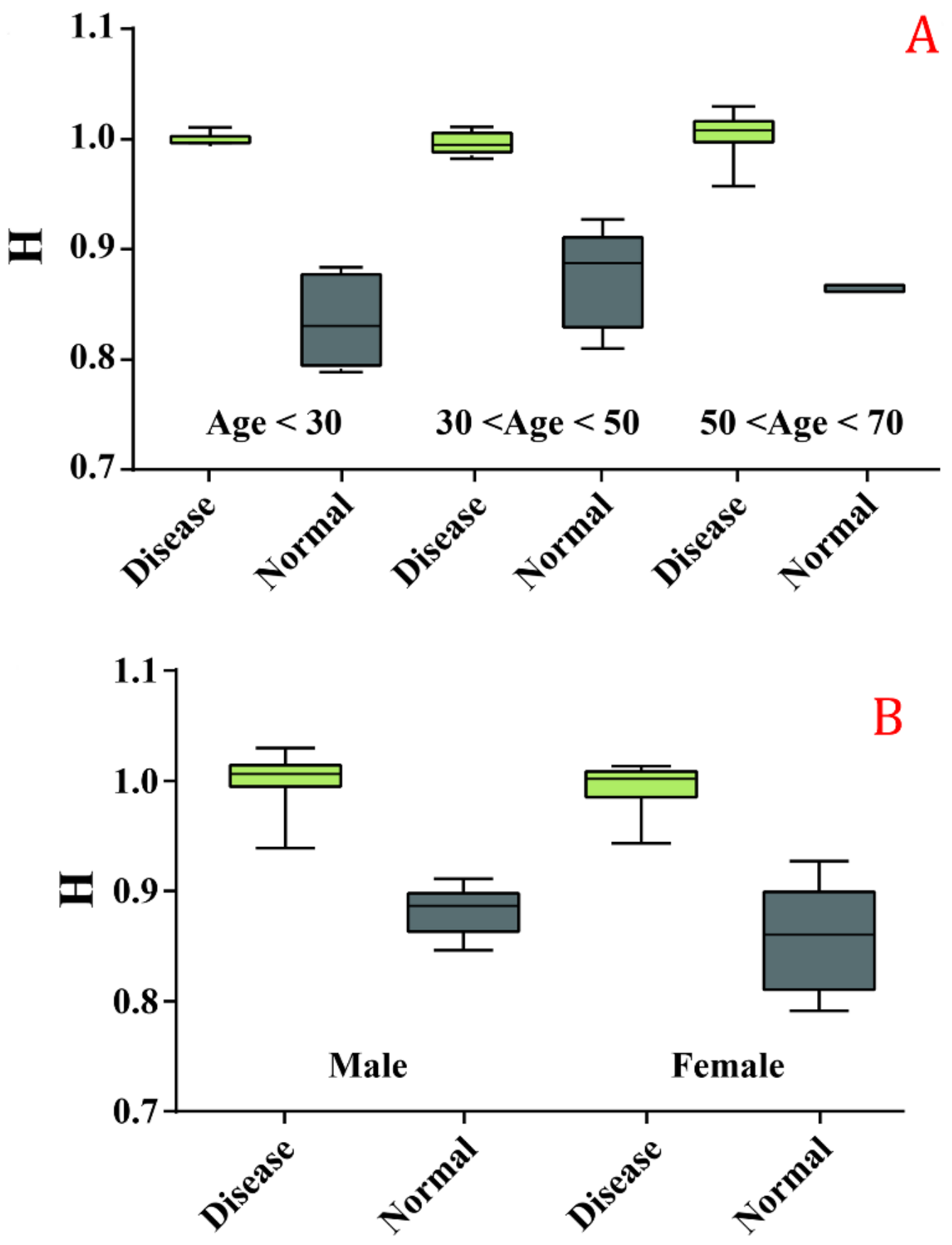

Figure 7: Box and whisker plot for H of all 1st IMFs of normal and diseased patients according to (A) age grouping and (B) gender grouping

The experimental result can also be enlightened with the reference of the patients' gender and age grouping. Here, such distribution will lead us to assess whether the H (of only 1$^{st}$ IMFs) is evenly distributed in all the range and also obey a similar pattern in short-range classification. In this connection, the H of all the 1$^{st}$ IMFs of diseased and normal patients are classified according to their gender, and the same is tabulated in Table IV. The mean of the H of the diseased male patients is 0.9992 (median = 1.0061) whereas, the same for the normal male patients is 0.8816 (median = 0.8866). 0.9950 (1.0006) and 0.8547 (0.8644) are the mean (median) of the H corresponding to the diseased and normal female patients respectively. The discrete margin between the H of normal and diseased patients is also consistent with the gender grouping (Fig. 7A). On the other hand, Table V reflects the details of the H of all 1$^{st}$ IMFs based on the age grouping of the patients. The value of H of the patients aged below 30 corresponding to the disease and normal subjects exhibit mean values as 1.0013 (median = 0.9994) and 0.8552 (median = 0.8736) respectively. The mean (median) values of H of disease and normal cases for the patients having age between 30 and 50 are evaluated as 0.9965 (0.9950) and 0.8809 (0.8821) consecutively. The same for the patients with age ranging from

Table IV: Statistical measures of H of 1st IMFs of the normal and diseased patients' data according to gender grouping

| | Patient ID | Age | H of 1st IMF | Mean | Median | p-value | Patient ID | Age | H of 1st IMF | Mean | Median | p-value |
|---|---|---|---|---|---|---|---|---|---|---|---|---|
| | Male patients | | | | | | Female patients | | | | | |
| Disease data | 230 | 32 | 0.9885 | 0.9992 | 1.0061 | 0.0001 | 208 | 23 | 0.9994 | 0.9950 | 1.0006 | 0.000002 |
| | 203 | 43 | 0.9950 | | | | 106 | 24 | 0.9966 | | | |
| | 214 | 53 | 1.0203 | | | | 113 | 24 | 1.0077 | | | |
| | 112 | 54 | 1.0189 | | | | 212 | 32 | 0.9821 | | | |
| | 233 | 57 | 1.0110 | | | | 115 | 39 | 1.0033 | | | |
| | 205 | 59 | 1.0061 | | | | 111 | 47 | 1.0137 | | | |
| | 213 | 61 | 0.9714 | | | | 119 | 51 | 1.0161 | | | |
| | 209 | 62 | 0.9544 | | | | 234 | 56 | 1.0042 | | | |
| | 107 | 63 | 1.0303 | | | | 123 | 63 | 0.9869 | | | |
| | 109 | 64 | 1.0115 | | | | 114 | 72 | 1.0044 | | | |
| | 200 | 64 | 0.9975 | | | | 231 | 72 | 1.0115 | | | |
| | 217 | 65 | 1.0324 | | | | 105 | 73 | 1.0129 | | | |
| | 201 | 68 | 0.9965 | | | | 101 | 75 | 1.0018 | | | |
| | 202 | 68 | 1.0096 | | | | 228 | 80 | 0.9687 | | | |
| | 100 | 69 | 0.9630 | | | | 121 | 83 | 1.0124 | | | |
| | 117 | 69 | 0.9964 | | | | 222 | 84 | 0.9792 | | | |
| | 118 | 69 | 1.0202 | | | | 108 | 87 | 0.9779 | | | |
| | 124 | 77 | 1.0157 | | | | 220 | 87 | 0.9847 | | | |
| | 215 | 81 | 0.9361 | | | | 207 | 89 | 0.9405 | | | |
| | 221 | 83 | 1.0004 | | | | | | | | | |
| | 103 | NR[a] | 1.0072 | | | | | | | | | |
| Normal Data | 16773 | 26 | 0.8866 | 0.8816 | 0.8866 | | 16272 | 20 | 0.8605 | 0.8547 | 0.8644 | |
| | 16265 | 32 | 0.8876 | | | | 16795 | 20 | 0.7886 | | | |
| | 19093 | 34 | 0.8436 | | | | 16273 | 28 | 0.8002 | | | |
| | 16483 | 42 | 0.9138 | | | | 16786 | 32 | 0.8126 | | | |
| | 19090 | 45 | 0.8766 | | | | 17453 | 32 | 0.9185 | | | |
| | | | | | | | 16539 | 35 | 0.8261 | | | |
| | | | | | | | 19140 | 38 | 0.9022 | | | |
| | | | | | | | 16420 | 38 | 0.8073 | | | |
| | | | | | | | 19088 | 41 | 0.8915 | | | |
| | | | | | | | 17052 | 45 | 0.9297 | | | |
| | | | | | | | 19830 | 50 | 0.8644 | | | |

50 to 70 are reflected in the said table as 1.0026 (median = 1.0078) and 0.8644 respectively (Fig. 7B). However, the assessment on the same platform for the aged patients (> 70 years) is skipped due to lack of normal patients' corresponding data for that age boundary. Therefore, in this table the distinction between the H corresponding to the diseased and normal subjects is also encountered whatever the assessment is made according to the age of the patients. Here, the p-test values for any grouping assessment are also listed in the above-mentioned tables and those show the results to be accepted herewith. All the 1st IMFs falls in the category of persistence type behavior as H of the same are always greater than 0.5. However, all the H for the disease data (~1) are greater than that of the normal data. The fact refers that the 1st IMFs series of the disease data exhibit a stronger trend than that of the normal case (Qian and Rasheed, 2004).

[a] NR=Not recorded

Table V: Statistical measures of H of 1st IMFs of the normal and diseased patients' data according to age grouping

| Age group | | Patient ID | Age (M/F) | H of 1st IMF | Mean | Median | | Patient ID | Age (M/F) | H of 1st IMF | Mean | Median | p-value |
|---|---|---|---|---|---|---|---|---|---|---|---|---|---|
| Age < 30 | Disease Data | 208 | 23 (F) | 0.9994 | 1.0013 | 0.9994 | Normal data | 16272 | 20 (F) | 0.8605 | 0.8552 | 0.8736 | 0.005215 |
| | | 106 | 24 (F) | 0.9966 | | | | 16795 | 20 (F) | 0.7886 | | | |
| | | 113 | 24 (F) | 1.0077 | | | | 16773 | 26 (M) | 0.8866 | | | |
| | | | | | | | | 16273 | 28 (F) | 0.8002 | | | |
| 30 < Age < 50 | | 212 | 32 (F) | 0.9821 | 0.9965 | 0.9950 | | 16265 | 32 (M) | 0.8876 | 0.8809 | 0.88209868 | 0.000001 |
| | | 230 | 32 (M) | 0.9885 | | | | 16786 | 32 (F) | 0.8126 | | | |
| | | 115 | 39 (F) | 1.0033 | | | | 17453 | 32 (F) | 0.9185 | | | |
| | | 203 | 43 (M) | 0.9950 | | | | 19093 | 34 (M) | 0.8436 | | | |
| | | 111 | 47 (F) | 1.0137 | | | | 16539 | 35 (F) | 0.8261 | | | |
| | | | | | | | | 19140 | 38 (F) | 0.9022 | | | |
| | | | | | | | | 16420 | 38 (F) | 0.8073 | | | |
| | | | | | | | | 19088 | 41 (F) | 0.8915 | | | |
| | | | | | | | | 16483 | 42 (M) | 0.9138 | | | |
| | | | | | | | | 19090 | 45 (M) | 0.8766 | | | |
| | | | | | | | | 17052 | 45 (F) | 0.9297 | | | |
| 50 < Age <70 | | 119 | 51 (F) | 1.0161 | 1.0026 | 1.0078 | | 19830 | 50 (F) | 0.8644 | 0.8644 | 0.8644 | |
| | | 214 | 53 (M) | 1.0203 | | | | | | | | | |
| | | 112 | 54 (M) | 1.0189 | | | | | | | | | |
| | | 234 | 56 (F) | 1.0042 | | | | | | | | | |
| | | 233 | 57 (M) | 1.0110 | | | | | | | | | |
| | | 205 | 59 (M) | 1.0061 | | | | | | | | | |
| | | 213 | 61 (M) | 0.9714 | | | | | | | | | |
| | | 209 | 62 (M) | 0.9544 | | | | | | | | | |
| | | 107 | 63 (M) | 1.0303 | | | | | | | | | |
| | | 123 | 63 (F) | 0.9869 | | | | | | | | | |
| | | 109 | 64 (M) | 1.0115 | | | | | | | | | |
| | | 200 | 64(M) | 0.9975 | | | | | | | | | |
| | | 217 | 65 (M) | 1.0324 | | | | | | | | | |
| | | 201 | 68 (M) | 0.9965 | | | | | | | | | |
| | | 202 | 68 (M) | 1.0096 | | | | | | | | | |
| | | 100 | 69 M) | 0.9630 | | | | | | | | | |
| | | 117 | 69 (M) | 0.9964 | | | | | | | | | |
| | | 118 | 69 (M) | 1.0202 | | | | | | | | | |

## 4. Conclusion

In the present paper, the EMD technique was performed on the nonlinear and non-stationary ECG signals to diagnose cardiovascular disease (arrhythmia). The IMFs, which are the fundamentally decomposed signal of a whole ECG signal, were extracted for arrhythmia arrested patients and also for normal subjects, using the EMD technique. A comparative study based on the enumerated H of the IMFs was carried out to distinguish the arrhythmia patients from the normal ones. In this analysis process, a distinct boundary between H of normal and diseased patients was encountered which indicates the contrast in dynamics of the cardiac system between normal and diseased individuals. The study shows that a probability of 94.92% was expected towards the predictability of arrhythmia disease in a patient. However, this analysis is restricted to the particular diseased datasets. A systematic analysis could be carried out on other cardiovascular disease data to ensure the predictability proportion. Moreover, the decision could be verified by means of applying other different nonlinear techniques such as wavelet analysis, multifractal analysis and detrended multifractal analysis, etc. This kind of analysis incorporated with machine learning might bring about a transformation in the current scenario of healthcare services.

## Acknowledgment

The authors are indebted to the National Institute of Technology Durgapur for providing institutional support for carrying out the research activities. The authors would also like to acknowledge Mr. Saroj Khutia, Ms. Kankana Seal and Ms. Suvashree Mukherjee for their sincere help to prepare the manuscript.

**Web references**